\documentclass[twocolumn]{aastex63}

\newcommand\kms{km s$^{-1}$}
\newcommand\masyr{mas yr$^{-1}$}
\newcommand\teff{$T_{eff}$}
\newcommand\logg{$\log g$}

\shorttitle{2M06464+0109: An EB of spotted, sub-solar twins}
\shortauthors{Miller et al.}

\graphicspath{{./}{figures/}}

\begin{document}

\title{Orbital and stellar parameters for 2M06464003+0109157: a double-lined eclipsing binary of spotted, sub-solar twins}

\correspondingauthor{Annaliese Miller}
\email{mille501@wwu.edu}

\author{Annaliese Miller}
\author{Marina Kounkel}
\author{Chase Boggio}
\author{Kevin Covey}
\affiliation{Western Washington University, 516 High St., Bellingham, WA 98225, USA}

\author{Adrian M. Price-Whelan}
\affiliation{Center for Computational Astrophysics, Flatiron Institute, New York, NY, 10010, USA}

\begin{abstract}

We calculated physical and orbital properties for 2M06464003+0109157, a 1.06 day eclipsing double lined spectroscopic binary. We modelled the system's ASAS-SN and TESS light curves, measuring the system’s inclination and radii of each component. Extracting radial velocities for each component from 6 SDSS/APOGEE spectra, we measured the system's mass ratio and performed a full orbital fit. Our analysis indicates that 2M06464003+0109157 has components with nearly equal masses ($m_1/m_2 =0.99 \pm 0.01$; $M_1{_,}{_2} = ~0.57 \pm 0.015$ M$_\sun$) and comparable radii ($R_1 = ~ 0.66 \pm 0.05$ R$_\sun$, $R_2 = ~ 0.57 \pm 0.06$ R$_\sun$). The solution required two star spots to incorporate the out of eclipse variation that is seen in the light curve. We report our full characterization of this system, and prospects for similar analyses using survey data to measure precise physical and orbital properties for EBs.

\end{abstract}


\section{Introduction} \label{sec:intro}

Eclipsing binaries (EBs) are extremely valuable objects, as they provide an opportunity to test stellar evolutionary models. Through characterizing a light curve of an EB, it is possible to constrain not only the inclination and eccentricity of an orbit, but also measure the radius of both stars (as a fraction of their semi-major axis). Furthermore, if complementary high resolution multi-epoch spectra are available, and both stars have a sufficiently high luminosity ratio and large enough separation in their respective radial velocities (RVs), the system may be detectable as a double-lined spectroscopic binary (SB2). In this case, from analyzing the RVs, it is possible to solve the full orbit and determine masses and true radii of the individual stars. Thus, EBs that are also SB2s can help to better constrain the mass-radius relation. There have been previous studies of the mass-radius relation on some stars in the past \citep[e.g.,][]{demircan1991,feiden2012,han2019,southworth2015}, however the number of such systems is still relatively low, particularly on the low mass end, as identification and analysis of EBs and SB2s require numerous photometric and spectroscopic observations spread over a large baseline.

In recent years, large surveys have made it possible to increase the sample of EBs with known masses and radii. In this study, we analyze a specific EB system to validate a workflow for deriving stellar \& orbital parameters using only serendipitous survey measurements, without any dedicated follow-up observations. By analyzing All-Sky Automated Survey for Supernovae (ASAS-SN) photometric monitoring of sources that had also been observed by Apache Point Observatory Galactic Evolution Experiment (APOGEE), \citet{pawlak2019} identified 430 EBs and reported their period estimates. We examined these systems using routines previously developed to identify and analyze SB2s in the APOGEE dataset \citep{kounkel2019}. In total we identified 53 that had a clear and well-resolved presence of both the primary and the secondary in the cross-correlation function of the individual spectral visits. Unfortunately, for most of these EBs, only a few epochs of APOGEE spectra are available, making it difficult to independently reconstruct a full orbit.

In this pilot study we characterize the properties of one system in particular, 2M06464003+0109157, hereafter 2M0646. This system is an EA type eclipsing binary with a 1.06493 day period \citep{pawlak2019}. At the time of the analysis, this particular system was a detached binary with the largest number of RV epochs to construct a full orbit with clear phasing of RVs to the eclipsing period.

2M0646 has kinematics and metallicity that are consistent with the stellar population typically identified as the Galaxy's Thick Disk component. Located just over 100 pc away, it has very high proper motion, in excess of 250 \masyr, or over 120 \kms. When combined with the system's center of mass velocity ($\sim$67 km/s; see section 4.2) and distance, these proper motions imply UVW velocities (U$_{lsr} = -19.2$ \kms; V$_{lsr}=-130.3$ \kms; W$_{lsr}=-20.1$ \kms) that are consistent with thick disk membership \citep[see Toomre diagrams in, e.g.][]{bensby2005,reddy2006,nissen2009}. Furthermore, 2M0646 has been reported to have low metallicity \citep[Fe/H $\sim$ -0.6 dex,][]{ahumada2020}. Both of these point to the system likely being quite old, with the age of $\sim$10 Gyr \citep[e.g.,][]{kilic2017}.

Although originally identified as an EB using the ASAS-SN data, 2M0646 has now been observed by TESS, allowing for a more pristine light curve analysis. The system was also observed in 6 epochs of the APOGEE spectra, with individual epochs well separated in the phase space for a full orbital fitting, making it a suitable target for mass and radius determination of both stars in the system.

In Section \ref{sec:identification} we describe the data available for 2M0646, both spectra and light curves. In Section \ref{sec:model} we perform fitting of all the available data, including light curves and the measured radial velocities to determine masses and radii of each star. Finally, we conclude in Section \ref{sec:discussion} by comparing the properties of this system to current mass-radius relations.

\section{Identification and Observations} \label{sec:identification}
\subsection{APOGEE spectra and RVs}
The APOGEE spectrograph \citep{wilson2010, wilson2012, majewski2017} covers the wavelength range of the H band (1.51-1.7 $\micron$), with a spectral resolution of R$\sim$22,500. 2M0646 has been observed by APOGEE over the course of 6 epochs, from HJD of 2456312 to 2456673, with distinct spectral reductions in the 14th and 16th data releases (DR14 and DR16) of the Sloan Digital Sky Survey \citep{blanton2017,abolfathi2018,ahumada2020}. \citet{olney2020} report parameters for 2M0646 of Teff =3870$\pm$26 K and $\log g$ = 4.416$\pm$0.027 dex, consistent with a low mass main sequence dwarf. Using the DR14 reduction of the data, 2M0646 is reported to have a metallicity of [Fe/H] = -0.302$\pm$0.016 dex; however, it drops to -0.6 using the DR16 reduction. We should note, that all of these pipelines are assuming a single star. It is difficult to estimate how an SB2 should affect the stellar parameters, though \citet{olney2020} note that they do not appear to systematically affect \teff\ or \logg\ distribution compared to single stars in a single age populations. Thus, most likely, they represent the parameters of the flux weighted mean of the two stars, which should usually gravitate to that of the primary.

In examining the cross-correlation function (CCF) of the system’s APOGEE spectra \citep{nidever2015}, we found that there was a signature of two strong components, indicating that this system is a double lined spectroscopic binary. In the DR14 data, components can be resolved in 4 out of 6 epochs, with the other two epochs unresolved due to the closeness of the velocities of the two stars. In the DR16 reduction, both components are resolved in all 6 epochs. The velocities of these components were extracted by fitting a Gaussian profile to the CCF using GaussPy \citep{lindner2015}. These velocities extracted from each reduction are presented in Table 1; when resolved in both data releases, individual component velocities are typically consistent to within 1 sigma, though the velocities measured at MJD=56316.7579 do differ by $\sim$2 sigma in each component. At the two epochs where the two components are not resolved in the DR14 CCFs, we report the velocity measured from the system’s unresolved CCF in the $v_1$ column, but note this quantity may better represent the system’s bulk heliocentric velocity than the velocity of any individual component at this epoch. 

We report velocities measured from both reductions, and we use DR16 reduction for the orbital fit (discussed in Section \ref{sec:full_fit}).

\begin{deluxetable*}{ccccc}
\tablecaption{Radial velocities for 2M0646, measured from APOGEE spectra}
\tablehead{
\colhead{Date} &
\colhead{$v_1$ (DR14)} &
\colhead{$v_2$ (DR14)} &
\colhead{$v_1$ (DR16)} &
\colhead{$v_2$ (DR16)} \\
\colhead{MJD} &
\colhead{\kms} &
\colhead{\kms} &
\colhead{\kms} &
\colhead{\kms}
}
\startdata
56312.7647605 &17.6$\pm$4.57 &114.5$\pm$6.4 &17.4$\pm$2.9 &110.8$\pm$4.4\\
56313.7623219 & -15.2$\pm$3.0 & 145.4 $\pm$ 4.3& -17.6$\pm$3.4 &143.6$\pm$6.0 \\
56316.7578667 & -27.1$\pm$2.7 &162.3 $\pm$ 4.2 &-30.5$\pm$1.5 &168.4$\pm$3.3\\
56349.676885 & 16.2$\pm$6.01 & 114.9$\pm$9.4 & 18.2$\pm$ 3.7 & 114.6$\pm$5.7 \\
56350.6830319 & 65.2$\pm$2.6 & \textit{N/A} & 45.7$\pm$18.2 & 82.9$\pm$27.1 \\
56672.8065246 & 64.2$\pm$ 2.5& \textit{N/A} &  90.9$\pm$14.7 & 52.5$\pm$17.1 \\
\enddata
\end{deluxetable*}

\subsection{Light curves}

\begin{figure*}[t]
\epsscale{0.8}
    \plotone{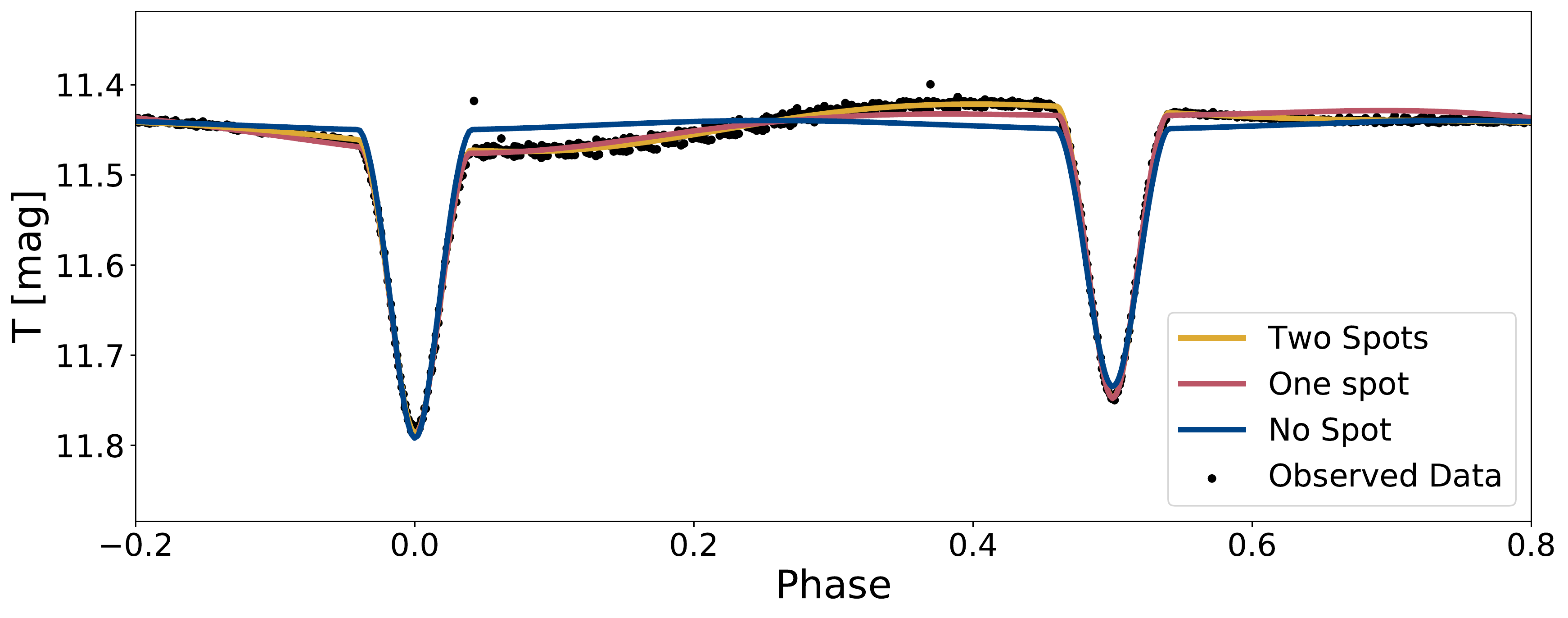}
    \caption{TESS light curve for 2M0646 (black dots) overlaid by multiple \texttt{ellc} models computed using maximum likelihood stellar \& orbital parameters with differing amounts of star spots.}
    \label{fig:fig5}
\end{figure*}

\begin{figure*}[t]
\epsscale{0.8}
    \plotone{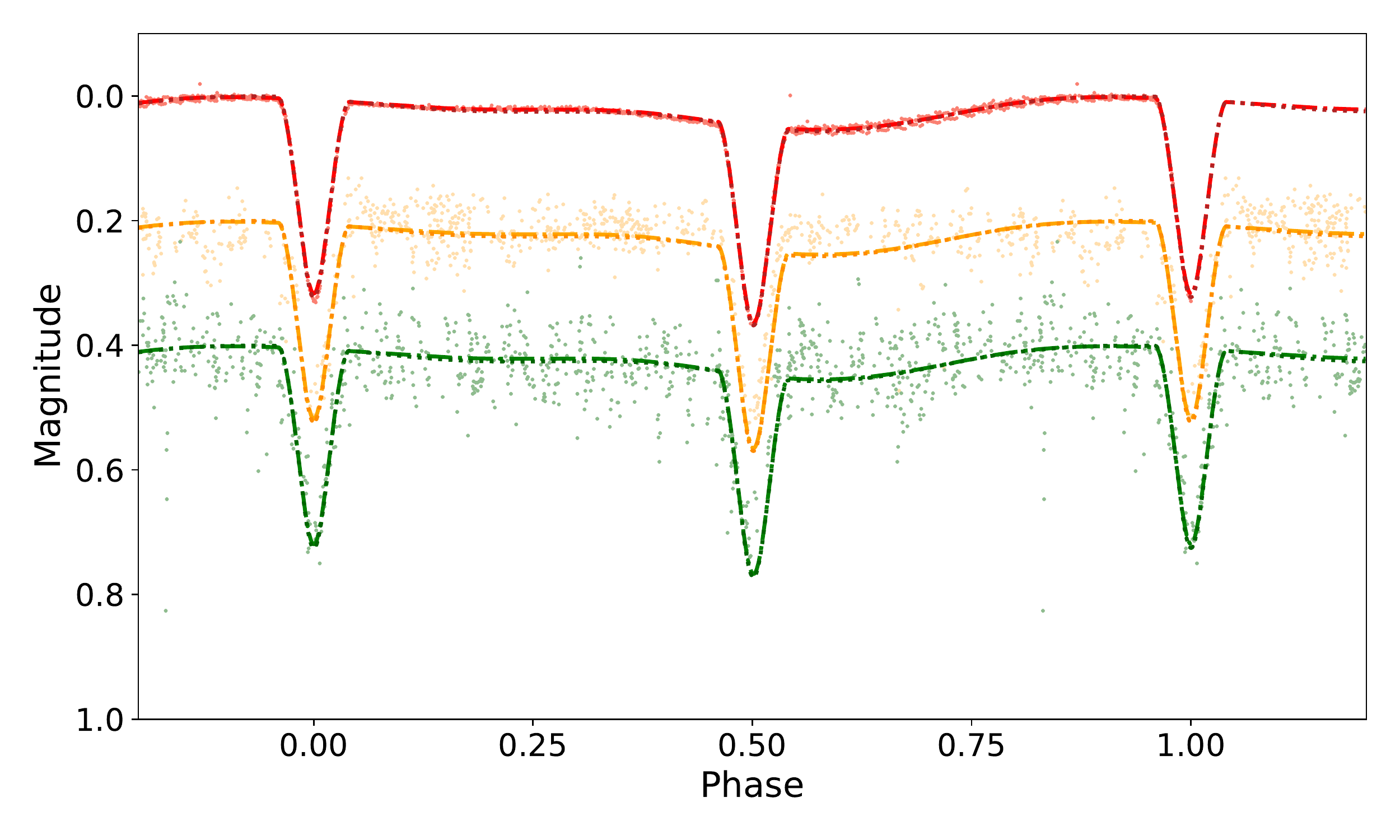}
    \caption{Final jointly fit LC models for the TESS (top), ASAS-SN V-band (middle) and ASAS-SN g-band (bottom) light curves.  Spot modulation is best fit in the TESS light curve, as out-of-eclipse datapoints in ASAS-SN light curves were intentionally down-weighted. The spot modulation signature that is fit to the TESS data appears consistent with the ASAS-SN g-band data, which is consistent with the relatively small temporal lag between the two datasets; the ASAS-SN V-band data was obtained somewhat earlier, and thus agrees less well with the spot modulation seen in the TESS data.}
    \label{fig:fig7}
\end{figure*}

2M0646 was identified as a $P=1.06493$ d eclipsing binary (EB) by \citet{pawlak2019} using light curves from the ASAS-SN survey \citep{kochanek2017}. ASAS-SN conducts nightly observations from multiple observing stations around the world. Each station contains four 14 cm aperture telephoto lenses, each equipped with a CCD camera that provides a field of view of 4.5 deg$^2$. ASAS-SN has observed 2M06464 over 1878 epochs from December 2014 to December 2020 at the time of analysis. Between 2014 and 2018, all epochs were observed in V-band. In 2018, ASAS-SN brought online their first g-band cameras and thus the observations were able to continue in both g-band and V-band. Even with the two different bands, the light curve was still sparsely populated. 

These data have been since superseded by the light curves from Transiting Exoplanet Survey Satellite \citep[TESS,][]{ricker2015}, which offer better cadencing and precision in the flux. TESS began its survey in July 2018. It is equipped with four CCD cameras that provide a simultaneous coverage of a 96x24 deg$^2$ area on the sky. Each sector of the sky is observed for a period of $\sim$27.4 days with a 30 minute cadence. 2M0646 is located in sector 6, producing a TESS light curve that spans the period from December 15th 2018 to January 7th 2019.

We used the \texttt{eleanor} pipeline \citep{feinstein2019} to download and extract the light curve of 2M0646 from the TESS full frame images archived in STScI’s MAST database. This pipeline provides an option for several different calibrations to detrend the light curves and reduce the systematics of the instrument. We adopt the light curve produced by the point spread function modelling, which performs well for brighter targets such as 2M0646.

The light curve is extracted with the photometry in the units of flux. However, the light curve fitting routine (Section \ref{sec:LCsol}) expects light curves to be magnitudes. We convert the flux to the T magnitude, ensuring that the average flux of the system is consistent with the magnitude reported in the TESS Input Catalog \citep[T=11.442 mag,][]{stassun2019}. The resulting light curve, phase-folded on a 1.0649d period, is shown in Figure \ref{fig:fig5}, which shows a significant improvement in precision and coverage compared to the ASAS-SN data in Figure \ref{fig:fig7}. In addition to better sampling of the eclipse themselves, the TESS light curve also reveals the presence of out-of-eclipse variations likely due to starspots.

\begin{deluxetable*}{lcccc}
\tablecaption{Spot Parameter Study -- System Parameters \label{tab:parameters}}  
\tablehead{
\colhead{} &
\colhead{} &
\colhead{One Spot} & 
\colhead{Two Spots} &
\colhead{} \\
\colhead{Parameter} & 
\colhead{units} &
\colhead{per component} &
\colhead{on Primary} &
\colhead{No Spots} }
\startdata 
\multicolumn5c{\textbf{Orbital Parameters}}  \\
Period & days & 1.065$\pm$0.019 &  1.065$\pm$0.008 &  1.065$\pm$0.009 \\
Inclination & $^{\circ}$ & 83.4$\pm$1.6 & 83.3$\pm$8.5 & 83.7$\pm$8.1 \\ 
\hline 
\multicolumn5c{\textbf{Stellar Parameters}}  \\
Primary Radius & (R/a) & 0.136$\pm$0.025 & 0.127$\pm$0.052 & 0.141 $\pm$ 0.059 \\ 
Secondary Radius & (R/a) & 0.136$\pm$0.024 & 0.143$\pm$0.067 & 0.136 $\pm$ 0.061 \\ \\
Surface Brightness Ratio & \textit{N/A} & 1.00$\pm$0.11 & 0.96$\pm$0.16 & 0.90$\pm$ 0.22 \\
%\hline \\
%\multicolumn5c{\textbf{Spot Parameters}}  \\
%Primary Spot Size & degrees ($^{\circ}$) & 34$\pm$16 & 37$\pm$15 & \nodata \\ 
% & & \nodata & 36$\pm$17 & \nodata \\ 
%%%%%%%Primary Spot Size (Spot 1) & degrees ($^{\circ}$) & 34$\pm$16 & 37$\pm$15 & \nodata \\ 
%%%%%~~~~~~~~~~~~~~~~~~~~~~~~ (Spot 2) & & \nodata & 36$\pm$17 & \nodata \\
%Secondary Spot Size & degrees ($^{\circ}$) & 31$\pm$15 & \nodata & \nodata \\ \\
%Primary Spot Longitude & degrees ($^{\circ}$) & 214$\pm$53 & 210$\pm$70 & \nodata \\
% & & & 205$\pm$92 & \\
%Secondary Spot Longitude & degrees ($^{\circ}$) & 210$\pm$140 & \nodata & \nodata \\ \\
%Primary Spot Latitude & degrees ($^{\circ}$) & 52$\pm$25 & -35$\pm$31 %& \nodata \\
% & & & -1$\pm$38 & \\
%Secondary Spot Latitude & degrees ($^{\circ}$) & 35$\pm$31 & \nodata & \nodata \\ \\
%Primary Spot Contrast & \textit{N/A} & 0.68$\pm$0.25 & 0.72$\pm$0.28 & \nodata \\
% & & & 0.84$\pm$0.31 & \\
%Secondary Spot Contrast & \textit{N/A} & 0.70$\pm$0.24 & \nodata & \nodata \\
\enddata
\end{deluxetable*}

%\vspace{-2cm}
\begin{deluxetable*}{lcccc}
 \tablecaption{Spot Parameter Study -- Starspot Parameters \label{tab:spots}}  
\tablehead{
\colhead{} &
\colhead{Spot} &
\colhead{} & 
\colhead{} &
\colhead{} \\
\colhead{} &
\colhead{Size} &
\colhead{Longitude} & 
\colhead{Latitude} &
\colhead{Spot} \\
\colhead{Model} & 
\colhead{degrees ($^{\circ}$)} &
\colhead{degrees ($^{\circ}$)} &
\colhead{degrees ($^{\circ}$)} &
\colhead{Contrast} }
\startdata
\multicolumn5c{\textbf{One Spot Per Component}} \\
Primary Spot & 34$\pm$16 & 214$\pm$53 &  52$\pm$25 &  0.68$\pm$0.25 \\
Secondary Spot & 31$\pm$15 & 210$\pm$140 &  35$\pm$31 &  0.70$\pm$0.24 \\ 
\hline 
\multicolumn5c{\textbf{Two Spots on Primary}}  \\
Sub-equatorial Spot & 37$\pm$15 & 210$\pm$70 & -35$\pm$31 & 0.72 $\pm$ 0.28 \\ 
Equatorial Spot & 36$\pm$17 & 205$\pm$92 & -1$\pm$38 & 0.84 $\pm$ 0.31
\enddata
\end{deluxetable*}

\section{Measuring Stellar and Orbital Parameters from LC \& RV fits}\label{sec:model}

\subsection{Light curve solutions: Degeneracies due to starspots} \label{sec:LCsol}

\texttt{ellc} \citep{maxted2016} is a detached binary star modeler that is designed to allow a Markov Chain Monte Carlo (MCMC) sampling mode. We used \texttt{ellc}, as driven by the \texttt{emcee} MCMC package \citet{foreman-mackey2013} to fit 2M0646’s light curve. Before performing a joint fit of the ASAS-SN, TESS, and APOGEE observations of this system (see next section), we conducted a parameter study to identify the sensitivity of the model results to the number and location of the starspots adopted in the fitting procedure.  As the spot modulation is more precisely measured, and less temporally blurred, in the TESS light curve, we focused this parameter study on fitting that light curve in isolation. 

For each trial in this parameter study, we initialized 200 MCMC walkers and computed chains for each walker with 20,000 steps. Starting with the broad distribution of parameters that were largely unconstrained, we allowed the model to converge and examined the outputs. Based on these solutions, we put increasingly  restrictive priors on the model parameters to exclude unphysical solutions (e.g., negative periods, face-on inclination, unrealistic ratio of the two radii, etc.), without cutting into the physically plausible parameter space to which most of the chains had converged. 

This process identified a core set of well motivated priors for the system's stellar and orbital parameters, which we used as consistent initial conditions while varying the number and location of spots in the model.  Specifically, we set a Gaussian prior on the period, centered at 1.06 days with a width of 0.025 days. The stellar radii are expressed by \texttt{ellc} as $R/a$, where $R$ is the physical radius of a given star, and $a$ is the semi-major axis of the system, thus the radius can be inferred from the width of the system’s eclipses in phase space. We set a Gaussian prior on $R/a$ centered on $R/a$ $\sim$ 0.13 with a width of $\pm$0.06. The inclination of the system was set at 83$\pm$15 degrees, and the surface brightness ratio as 0.95$\pm$0.25. Eccentricity was set as zero. The time of passage of the eclipse $T_0$ was restricted with a uniform prior from 0 to 1.1 (to prevent the solution wrapping around the phase), and was set centered at 0.75 days. 

Using these priors, we then attempted to fit the TESS light curve with models assuming no star spots, one spot (on the system's primary), one spot per star, and two spots on the primary component.  The best fitting model from each of these spot configurations is shown in Figure \ref{fig:fig5}, and system and spot parameters are given in Tables \ref{tab:parameters} and \ref{tab:spots}, respectively.  As Figure \ref{fig:fig5} shows, the best fit from the no-spot model is unable to reproduce the system's out-of-eclipse variations. Adding a single spot to the model significantly improves the fit, but still leaves systematic residuals in phase. 

Ultimately, we find two spots are necessary to fully fit the out-of-eclipse light curve. In the two spots, one-per-component model, the spots were initialized with a size of $40\pm20$ degrees, a longitude ($l$) of $180\pm180$ degrees, a latitude ($b$) of $30\pm30$ degrees, and a brightness factor of $0.7\pm0.3$. At the final initialization of the system, only one uniform prior was placed on $l$ from 0 to 360 degrees to keep it physically plausible.  However, the model is not sensitive to the distribution of spots - the fit is comparable if both spots are located on the same stars versus on one star each. Similarly, the model is not sensitive to discriminating between dark spots and bright faculae. The resulting parameters across different models are shown in Table 2. The preferred spot sizes ($\sim35$ degrees) and contrasts ($\sim70$\%) are consistent across both dual-spot configurations. 

Although the average $l$ and $b$ show a wide distribution in Table \ref{tab:spots}, this average is a superposition of various discrete locations that are equally favored. If faculae are used, they tend to be positioned facing towards the other star in the binary, whereas darker spots tend to be located on the opposite side. We interpret this as two stars that are tidally locked irradiating each other’s photosphere, and consider the model where both spots are equally distributed across the stars as most likely.

The addition of the star spots to the solution does shrink the radii of the individual stars and increase the surface brightness ratio compared to the spotless model. This is due to \texttt{ellc} suppressing the average brightness of the spotted primary. Both single and double spot models produce a comparable effect. The spotted models are more consistent with the near-equal mass ratio system (see Section \ref{sec:full_fit}). Although both solutions that have two star spots fit the LC well, \replaced{whether on both on the primary or on each component}{whether both of the spots are on the primary, or there is one on each of the stars}, there is an ~6\% radii difference between the two which represents the systematic uncertainty in the solution.  We chose, in our final fit, to adopt the 'one-spot-per-component' model, as this seemed most consistent with the underlying symmetry in the system's parameters/mass ratio.

\subsection{Full orbital fits: joint light curve and radial velocity fits}\label{sec:full_fit}

\begin{deluxetable}{cccc}
\tablecaption{Priors \& Initialization of stellar + orbital parameters}\label{tab:priors}
\tablehead{
\colhead{} &
\colhead{} & 
\colhead{Initial} &
\colhead{Initial} \\
\colhead{Parameter} &
\colhead{Priors} & 
\colhead{Median} &
\colhead{$\sigma$}
}
\startdata
$P$ (days) & NA & 1.06494 & 1e-6 \\
$T_0$ & NA & 2458472.537 & 0.001 \\
$\gamma$ (\kms) & -200 -- 200 & 67 & 2 \\
$K_1$ (\kms) & NA & 107 & 2.5 \\
$q$ & 0 -- 1 & 0.98 & 0.01 \\
$e$ & 0 -- 1 & 0.01 & 0.02 \\
$i$ (deg) & 0 -- 90 &  83.4  &  2.0 \\
$\omega$ (radians) & 0 -- 2$\pi$ & 4.97 & 0.52 \\
Surface brightness ratio & $>$0 & 0.94  & 0.02 \\
\hline
R$_1$/a & 0 -- 1 & 0.145 & 0.02 \\
R$_2$/a & 0 -- 1 & 0.13 & 0.02 \\
\hline
Primary spot size (deg) & 0 -- 90 & 31 & 10 \\
Spot longitude (deg) & 0 -- 360 & 207 & 30 \\
Spot latitude (deg) & -90 -- 90 & 40 & 30 \\
Spot contrast ratio & 0 -- 1 & 0.65 & 0.2 \\
\hline
Secondary spot size (deg) & 0 -- 90 & 18 & 10 \\
Spot longitude (deg) & 0 -- 360 & 260 & 30 \\
Spot latitude (deg) & -90 -- 90 & 22 & 30 \\
Spot contrast ratio & 0 -- 1 & 0.65 & 0.2 \\
\enddata
\end{deluxetable}

Initially, we explored extracting stellar and orbital parameters by using \texttt{ellc} to fit the system's light curve and \textit{The Joker} \citep{price-whelan2017,price-whelan2018} to fit the system's radial velocity measurements.  Given the disjoint epochs of the TESS light curve and the APOGEE radial velocities, and the longer but sparser ASAS-SN monitoring, however, fitting the light curves and radial velocities separately led to non-negligible uncertainties in the system's period and T$_0$ ($\sim$0.02 days).  

To best constrain the system's orbital period and ephemeris, we utilized \texttt{ellc}'s light curve and radial velocity modelling capabilities to jointly fit the system's TESS \& ASAS-SN (V- and g-band) light curves, as well as the APOGEE radial velocities.  In this joint fit, we utilize the ASAS-SN light curves mainly for the constraints their in-eclipse datapoints provide for the system's period and ephemeris; as 2M0646's starspots may evolve substantially over the baseline of the ASAS-SN monitoring, we artificially down-weight the ASAS-SN out-of-eclipse data points (which we define as any datapoint with phase between 0.05 -- 0.45, and 0.55 -- 0.95) by inflating their errors by a factor of 100.  We then use \texttt{emcee} to identify 2M0646's maximum likelihood stellar and orbital parameters by using \texttt{ellc} to compute self-consistent light curve and radial velocity solutions for a given set of model parameters, and then compute a joint chi square value from the residuals of the fits to the TESS, ASAS-SN and APOGEE data.  Given the relatively small number of radial velocity datapoints, and their orthogonal information content relative to the light curve, we apply a 20x multiple to the APOGEE residuals in computing the joint chi square value. In computing the light curve models, we adopt limb darkening coefficients computed by \citet{Claret2011} for the V- and g-bands (for the ASAS-SN light curves), and for a white-light/Kepler bandpass (for the TESS light curve).
In this joint fit, we applied priors sparingly, primarily to limit the solution to physically plausible solutions (i.e, restricting eccentricity to 0 $< e <$ 1, the mass ratio to 0 $< q <$ 1, etc.). We initialized the chain with Gaussian distributions around parameters which were found to well fit the light curve in the parameter study described earlier.  Specifically, we initialized the walkers using the mean values and standard deviations listed in Table \ref{tab:priors}.

\begin{figure}
    \centering
    \epsscale{1.25}
    \plotone{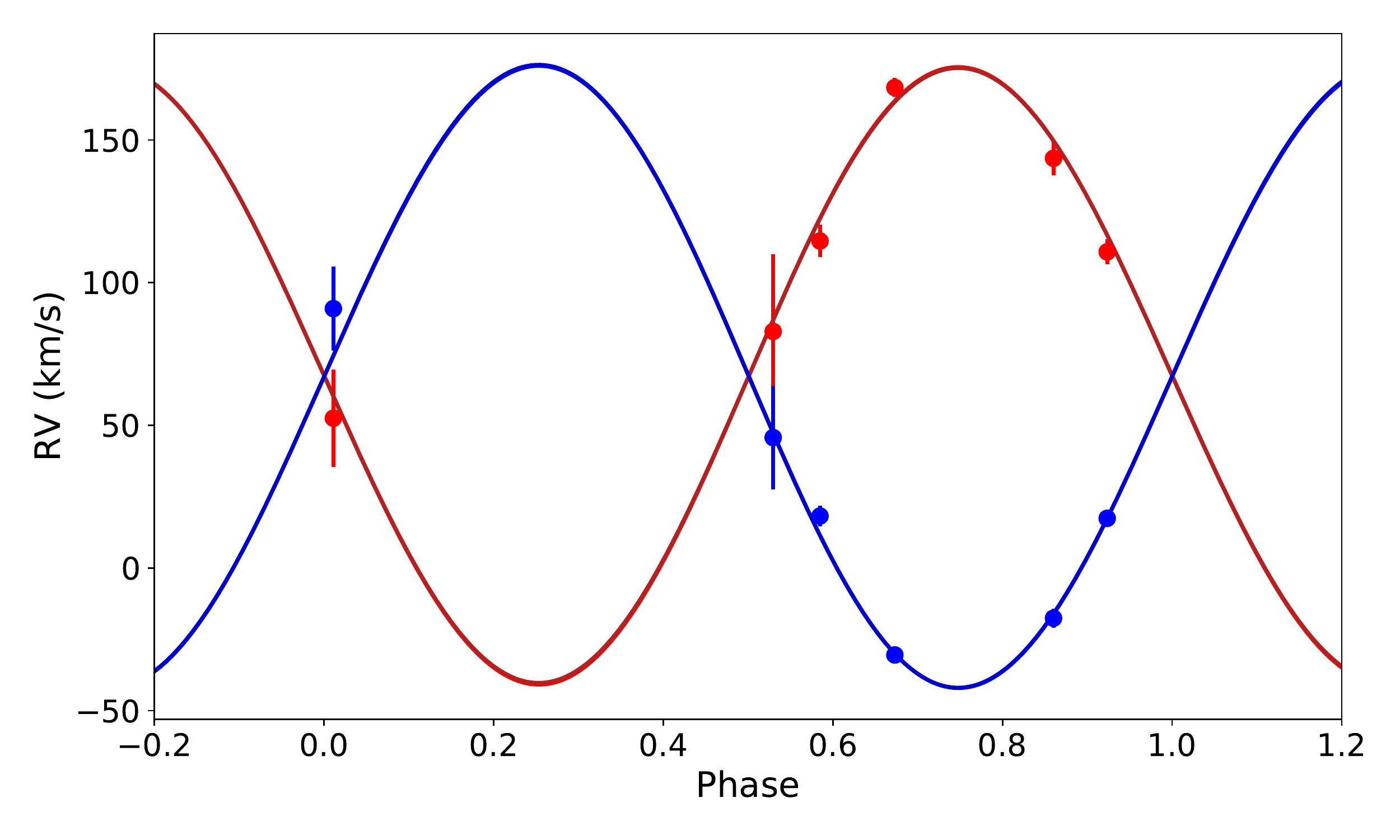}
    \plotone{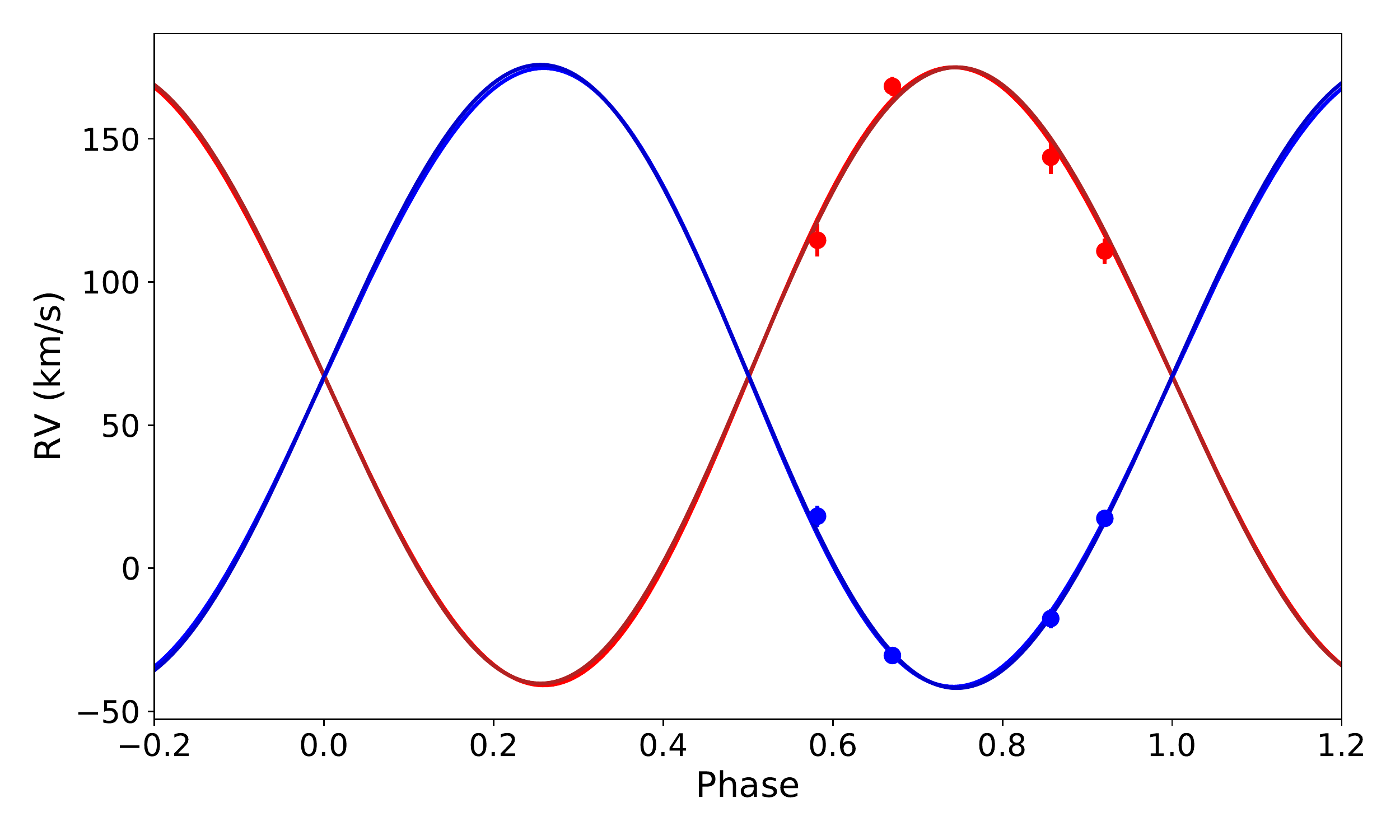}
    \caption{RV vs. time plot, showing individual sample orbits fit by \texttt{ellc} (blue \& red lines) to the velocities measured for 2M0646's primary (red dots) and secondary (blue dots) respectively. Two different models are shown, the top one has included the two RV points close to the eclipses, the bottom one has excluded them from the fit.}
    \label{fig:fig8}
\end{figure}

The remaining parameters that are customizable by \texttt{ellc} (e.g., third light, irradiation, apsidal motion rate, asynchronous rotation factor, surface gravity, $[M/H]$, reflection coefficient, external noise, $T_{\rm{exp}}$) were not modified from their default values, as appropriate for a detached system of solar-metallicity main sequence dwarfs, or set to zero in the initial run.

We initialize a model with 15 walkers, each of which takes 5000 steps. We used the MCMC chain to estimate parameters for $r_1$, $r_2$, surface brightness ratio, inclination, $T_0$, and period. Before measuring parameters from the chain, we reject the first 10\% of the chain to ensure that it has fully burned in (i.e., become relatively independent of where it was initialized). We then calculate the median and standard deviation of each parameter in the remainder of the chain, and adopt these values to characterize the system.

To investigate the effect of the two RV pairs of points near the eclipses (using DR16 data) on the derived orbit, we repeat the joint \texttt{ellc} fit and exclude them. The two RV fits are shown in Figure \ref{fig:fig8}, and have negligible difference, either visually in examining the fitted orbit, or in the derived parameters.

\begin{figure}
    \centering
    \epsscale{1.2}
    \plotone{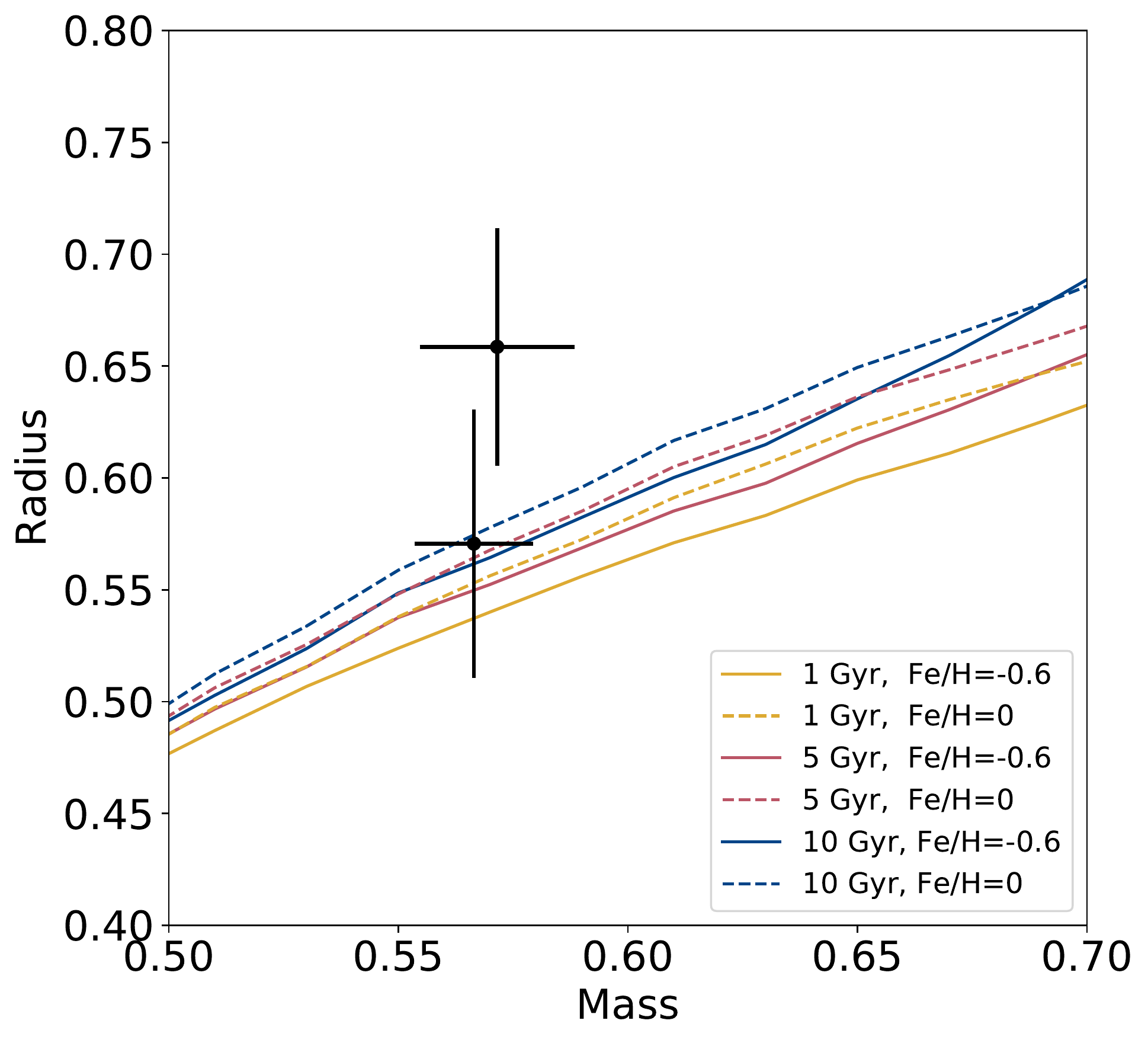}
    \caption{Mass and radius relation of the two components of 2M0646 compared against the PARSEC isochrones of different ages and metallicities.}
    \label{fig:fig9}
\end{figure}

\begin{figure*}
    %\epsscale{0.8}
    \plotone{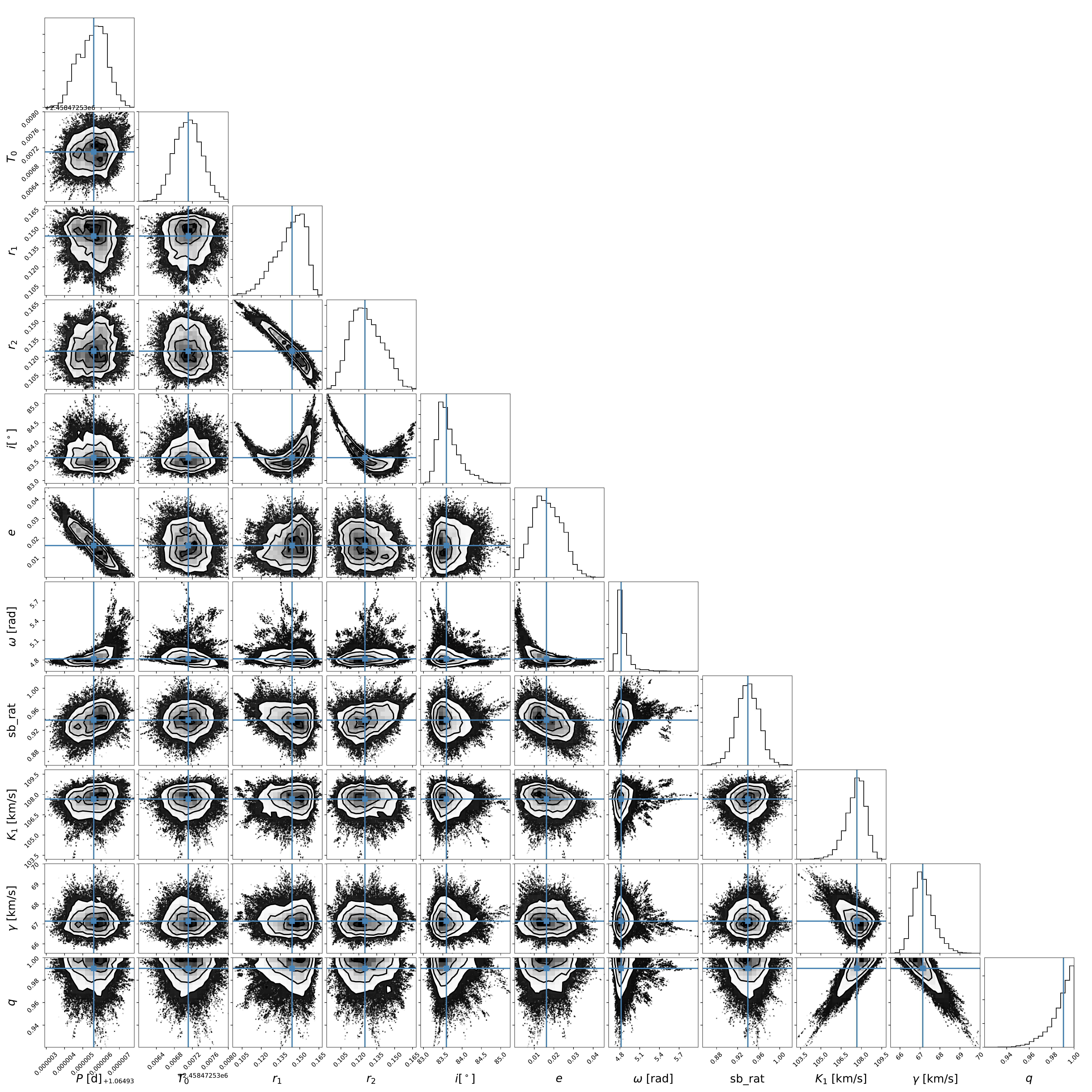}
    \caption{Corner plot showing distribution of stellar and orbital parameters after final iteration of two spot model (with one spot per component). The parameters are labelled as, orbital period ($P$, in days), time of passage of periastron ($T_0$, julian date), radius of the primary and the secondary ($r_1/r_2$ in units of the system’s semi-major axis), orbital inclination ($i$, in degrees), eccentricity of the system ($e$), argument of periapsis ($\omega$, in radians), surface brightness ratio (sb\_rat), velocity semi-amplitude of the primary ($K_1$, in \kms), the center of mass velocity ($\gamma$, in \kms), and mass ratio ($q$).}
    \label{fig:fig4}
\end{figure*}

\begin{figure*}
    \centering
    %\epsscale{0.8}
    \plotone{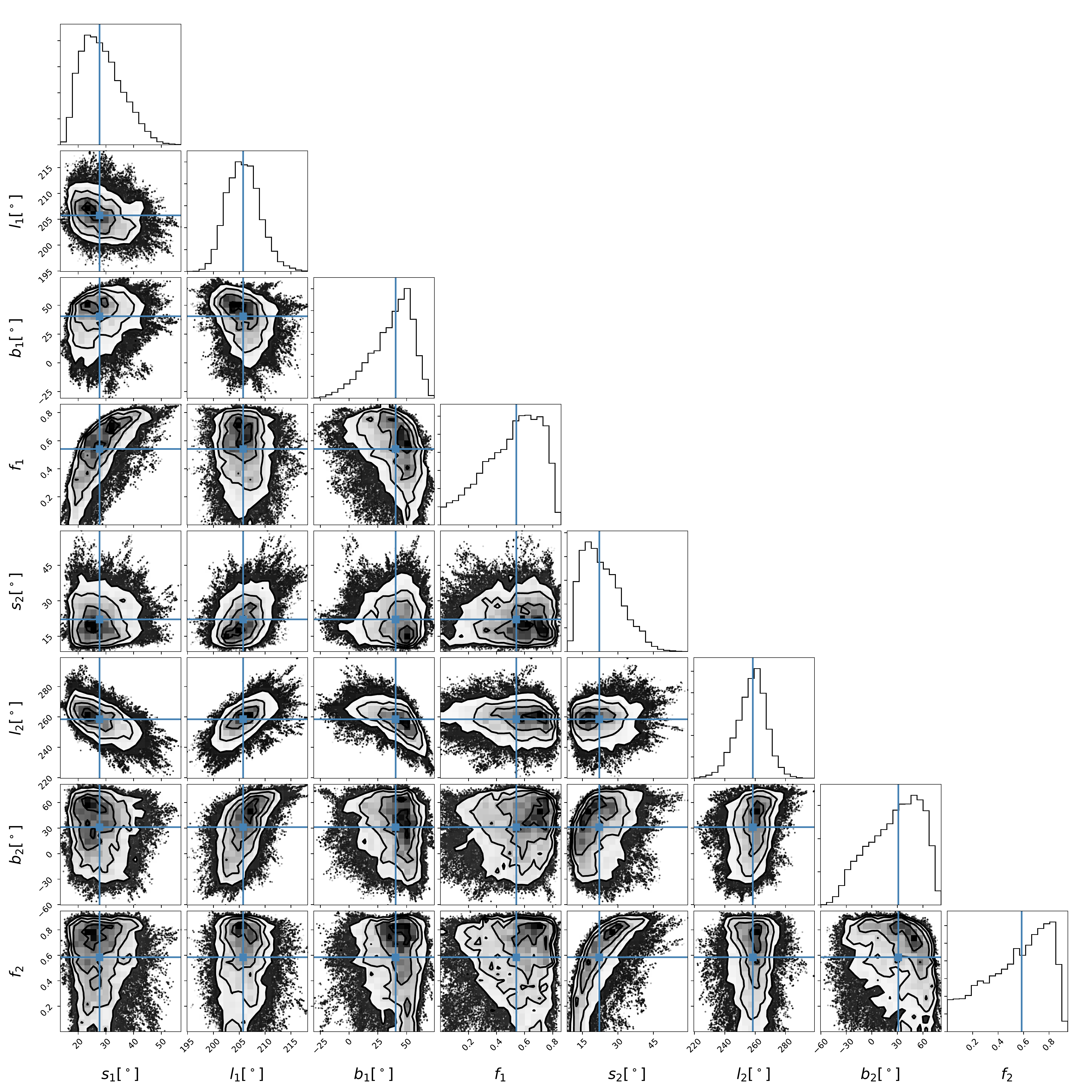}
    \caption{Corner plot showing distribution of starspot properties in final two-spot \texttt{ellc} model (w/ one spot per component). The labelled parameters are the size of the spots on both primary and the secondary ($s_1/s_2$), in degrees, the surface longitude of the spots on each star ($l_1/l_2$, in degrees), surface latitude ($b_1/b_2$, in degrees), and the respective ratio of the contrast of the spot relative to the photosphere ($f_1/f_2$).}
    \label{fig:fig6}
\end{figure*}

Best fitting parameters for the model containing all the data are given in Table \ref{tab:best_fit}. The derived masses and radii are show in Figure \ref{fig:fig9}. The full corner plots of star+orbit and spot parameters are shown in Figures \ref{fig:fig4} and \ref{fig:fig6} respectively.  To convert the radii from units of semi-major axis to solar radii, we used a variation of Kepler’s 3rd Law to calculate the semi-major axis of the system. We then multiply the resultant semi-major axis by the relative radii ($r/a$) produced by \texttt{ellc}. Performing this conversion and propagating the errors in $r/a$, $P$, $i$, and $v_1+v_2$, we find $R_1 = 0.66 \pm 0.05$ R$_\sun$ and $R_2 = 0.57 \pm 0.06$ R$_\sun$.

\begin{deluxetable}{cc}
\tablecaption{Final physical parameters for two spot solution}\label{tab:best_fit}
\tablehead{
\colhead{Parameter} &
\colhead{Value}
}
\startdata
$P$ (days) & 1.06493559$\pm$7.5e-07 \\
$\gamma$ (\kms) & 67.13$\pm$0.52 \\
$K_1$ (\kms) & 107.6$\pm$0.7 \\
$K_2$ (\kms) & 108.6$\pm$1.4 \\
$q$ & 0.99$\pm$0.01 \\
$e$ & 0.016 $\pm$0.007 \\
$i$ (deg) & 83.6  $\pm$  0.3 \\
$\omega$ (deg) & 276$\pm$6\\
Surface brightness ratio & 0.94  $\pm$  0.02 \\
$T_0$ & 2458472.5371  $\pm$  0.0003 \\
\hline
$M_1$ (M$_\sun$) & 0.571$\pm$0.017 \\
$M_2$ (M$_\sun$) & 0.565$\pm$0.013 \\
$R_1$ (R$_\sun$) & 0.659$\pm$0.053 \\
$R_2$ (R$_\sun$) & 0.570$\pm$0.060 \\
\hline
Primary spot size (deg) & 27.7 $\pm$ 7.2 \\
Spot longitude (deg) & 205.7 $\pm$3.2 \\
Spot latitude (deg) & 40 $\pm$18 \\
Spot contrast & 0.54 $\pm$0.20 \\
\hline
Secondary spot size (deg) & 22.1 $\pm$ 8.2 \\
Spot longitude (deg) & 258.5 $\pm$9.6 \\
Spot latitude (deg) & 31 $\pm$29 \\
Spot contrast & 0.58 $\pm$0.24 \\
\enddata
\end{deluxetable}

\section{Discussion and conclusions}\label{sec:discussion}

We have identified an eclipsing SB, 2M0646, and have performed a orbital and light curve fitting to measure masses and radii of the individual stars of the binary. We measure the properties of each component to be 0.57$\pm$0.02 M$_\sun$ and 0.66$\pm$0.05 R$_\sun$ for the primary, and 0.56$\pm$0.01 M$_\sun$ and 0.57$\pm$0.06 R$_\sun$ for the secondary. The secondary's derived properties are consistent within the $\sim$10\% errors with the mass-radius relations suggested by the PARSEC isochrones \citep{marigo2017}, but the primary appears over-inflated at the 1.5$\sigma$ level with respect to those isochrones' predictions. We note that there various degeneracies in radius determination with respect to e.g., inclination (Figure \ref{fig:fig4}), as well as spots coverage (Table \ref{tab:parameters}). Given the similarity of the masses, it is unlikely that there is a significant difference in the radii, and, indeed the individual models of various subsets of the data do suggest it. Thus, most likely, the radii for both stars are likely to be $\sim$0.6 $R_\odot$, i.e., consistent with the overlap in the 1$\sigma$ between the two measurements reported in Table \ref{tab:best_fit} and Figure \ref{fig:fig9}, with both stars moderately inflated in comparison to the isochrones.

Lower metallicity isochrones with older ages (which are more appropriate for thick-disk sources such as 2M0646) tend to have marginally larger radii at a given mass than the solar metallicity isochrones. This is likely insufficient to fully compensate for the over-inflation, furthermore, the differences between various isochrones are smaller than the uncertainties in our radius measurements. Additional monitoring, and more detailed abundance measurements, would provide a better basis for a precision test of the isochrones' accuracy for stars of non-solar age and metallicity.

For nearly a decade and a half precision measurements have identified that theoretical models often underpredict the empirically measured radii of low-mass stars.  The presence and magnitude of these inflated radii have been reported to correlate with metallicity \citep[e.g., ][]{Berger2006}, direct signatures of stellar activity \citep[e.g.,][]{Lopez-Morales2007}, and/or orbital/rotational velocities indicative of likely induced activity \citep{Kraus2011a, Somers2017}.  These observational results have spurred the development of theoretical models that seek to resolve this discrepancy by incorporating new physics into the calculation of isochrones and mass radius relations for low-mass stars, such as direct magnetic inhibition of convection \citep[][]{Feiden2014, MacDonald2017} and the reduced efficiency of radiative losses by spotted photospheres \citep{Somers2020}. The picture is muddled, however, by other recent observational studies which find either no evidence for inflated radii \citep[e.g., ][]{Han2017} , or no clear correlation with metallicity/rotation/activity \citep[e.g., ][]{Kesseli2018, Parsons2018, Morrell2019}.  

To the extent that our measurements do suggest that the system's primary may be inflated at the 1.5 sigma level relative to the predictions of theoretical isochrones, it is difficult to draw broader conclusions about the likely cause of that inflation.
Effects related to magnetic activity and/or sunspots are a plausible explanation: with clear spot modulation on a period that matches the system's short orbital period, the system appears tidally locked and a good candidate for long-lived magnetic activity with a rotationally synchronized dynamo.
Similarly, the two stars' close proximity and near-equal mass ratio suggests that irradiation from the companion will allow each component to affect it's neighbor's surface temperature and thermal structure, potentially inhibiting convection and producing an inflated radius.

In the future, additional photometric monitoring of 2M0646 to obtain color information may help to further diagnose the temperature contrast and location of spots.

Spectrophotometric monitoring of the system would provide even more information: a detailed map not only of the location and magnetic properties of the spots, but also an opportunity to perform detailed spectral fitting to test for compositional differences.  More broadly, more observations by large spectroscopic surveys such as SDSS that allow serendipitous detections of SB2s, combined with the new EB detections in the light curves that are now available across much of the sky with TESS, will provide additional systems with reliable masses and radii to conduct statistical tests of potential inflation mechanisms.

\software{\texttt{ellc} \citep{maxted2016}, \texttt{eleanor} \citep{feinstein2019}, The Joker \citep{price-whelan2017,price-whelan2018},\texttt{emcee} \citep{foreman-mackey2013}}

\section{Acknowledgements}
We thank the anonymous referee whose useful and constructive feedback improved the substance and presentation of our analysis. M.K. and K.C. acknowledge support provided by the NSF through grant AST-1449476; A. M. acknowledges additional support from the National Aeronautics and Space Administration, provided by the Washington NASA Space Grant Consortium and Chandra Award Number GO9-20006X issued by the Chandra X-ray Center, which is operated by the Smithsonian Astrophysical Observatory for and on behalf of the National Aeronautics Space Administration under contract NAS8-03060.

Funding for the Sloan Digital Sky 
Survey IV has been provided by the 
Alfred P. Sloan Foundation, the U.S. 
Department of Energy Office of 
Science, and the Participating 
Institutions. 

SDSS-IV acknowledges support and 
resources from the Center for High 
Performance Computing  at the 
University of Utah. The SDSS 
website is www.sdss.org.

SDSS-IV is managed by the 
Astrophysical Research Consortium 
for the Participating Institutions 
of the SDSS Collaboration including 
the Brazilian Participation Group, 
the Carnegie Institution for Science, 
Carnegie Mellon University, Center for 
Astrophysics | Harvard \& 
Smithsonian, the Chilean Participation 
Group, the French Participation Group, 
Instituto de Astrof\'isica de 
Canarias, The Johns Hopkins 
University, Kavli Institute for the 
Physics and Mathematics of the 
Universe (IPMU) / University of 
Tokyo, the Korean Participation Group, 
Lawrence Berkeley National Laboratory, 
Leibniz Institut f\"ur Astrophysik 
Potsdam (AIP),  Max-Planck-Institut 
f\"ur Astronomie (MPIA Heidelberg), 
Max-Planck-Institut f\"ur 
Astrophysik (MPA Garching), 
Max-Planck-Institut f\"ur 
Extraterrestrische Physik (MPE), 
National Astronomical Observatories of 
China, New Mexico State University, 
New York University, University of 
Notre Dame, Observat\'ario 
Nacional / MCTI, The Ohio State 
University, Pennsylvania State 
University, Shanghai 
Astronomical Observatory, United 
Kingdom Participation Group, 
Universidad Nacional Aut\'onoma 
de M\'exico, University of Arizona, 
University of Colorado Boulder, 
University of Oxford, University of 
Portsmouth, University of Utah, 
University of Virginia, University 
of Washington, University of 
Wisconsin, Vanderbilt University, 
and Yale University.

\bibliography{main.bbl}{}
\bibliographystyle{aasjournal}

\listofchanges

\end{document}